\documentclass{article}  
\usepackage{breckenridge}
\frompage{000} \topage{000}                                              
\def\rightmark{Do Ordinary Nuclei Contain Exotic States of Matter?} 
\def\leftmark{J. Arrington}
 
\title{Do Ordinary Nuclei Contain Exotic States of Matter?} 
\authors{ {John Arrington$^1$}\\[2.812mm]
{\normalsize \hspace*{-8pt}$^1$ Argonne National Laboratory, \\ 
Argonne, IL, 60439, USA\\[0.2ex] 
}}

\abstract{The strongly repulsive core of the short-range nucleon-nucleon
interaction leads to the existence of high-momentum nucleons in nuclei.
Inclusive electron scattering can be used to probe these high-momentum
nucleons and study the nature of the corresponding short-range correlations in
nuclei. With recent data from Jefferson Lab we have begun to map out the
strength of two-nucleon correlations in nuclei, while upcoming experiments
should allow us to isolate the presence of multi-nucleon correlations. In
addition to their importance in describing nuclear structure, these
configurations of correlated nucleons represent high density 'droplets' of
hadronic matter. As the density of hadronic matter increases there should be a
weakening of quark confinement, similar to the onset of deconfinement expected
at extremely high temperatures. While there have been hints of non-hadronic
structure in nuclei, future measurements will allow us to directly probe the
quark distributions of high density configurations in nuclei. A modified quark
structure in these closely packed nucleons would provide a clear signature of
exotic components to the structure of nuclei.}

\keyword{keywords here} 
\PACS{12.38.Mh,24.85.+p,25.30.Fj}
 
\begin{document}
 
\maketitle
\setcounter{page}{1}

\section{Inclusive Scattering from Nuclei at $x>1$}\label{sec:intro}

        Inclusive scattering from nuclei at $x>1$ has provided a great deal of
information about the high momentum components of the nuclear wave function.
Measurements at large momentum transfer, $q$, and small energy transfer,
$\nu$, are dominated by quasielastic scattering from single nucleons.  This
process is sensitive to the elastic electron-nucleon cross section and the
distribution of nucleons within the nucleus.  For a stationary nucleon,
elastic scattering corresponds to $x_{Bjorken} = Q^2 / 2 m_N \nu = 1$, where
$m_N$ is the mass of the nucleon and $Q^2 = q^2 - \nu^2$.  Nucleons with
a large initial momentum parallel (anti-parallel) to the exchange photon
occur at $x>1$ ($x<1$).  By selecting low energy transfers, corresponding
to $x>1$, there is little background from inelastic processes, and the
measurements are sensitive to the distribution of nucleons with large momenta.

These high momentum components come from short-range correlations (SRCs)
in the nucleus.  These correlations are an important part of nuclear structure
and are related to the short-range, strongly repulsive core of the N-N
interaction. Comparisons of inclusive scattering from heavy nuclei and
deuterium provided one of the first experimental signatures of short-range
range correlations in nuclei~\cite{ratiodata,ratiocalc}.  Above $x \approx
1.5$, the cross section behavior was identical for all nuclei measured,
yielding a plateau in the $A/D$ ratio and indicating that the same two-nucleon
correlations that generate the high-momentum nucleons in the deuteron are also
the dominant source of high-momentum nucleons in heavy nuclei. Subsequently,
short-range correlations in nuclei have been probed in a variety of
reactions including A(p,2p), A(e,e'p), A(e,e'NN).

\begin{figure}[htb]
\vspace*{-.4cm}
                 \insertplot{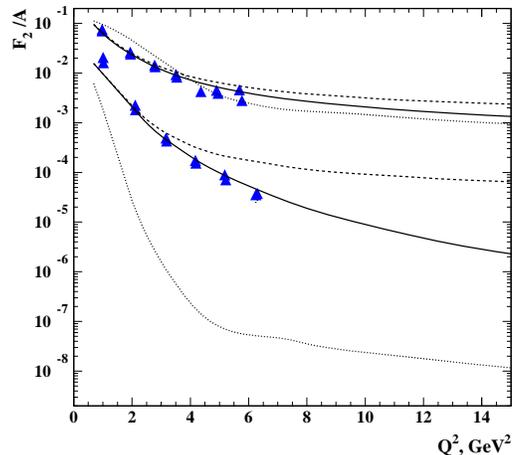}
\vspace*{-1.6cm}
\caption[]{Structure function per nucleon for iron~\cite{jlabexpts} from
JLab E89-008 compared to calculations without correlations (dotted lines),
including two-nucleon SRCs (solid lines), and multi-nucleon SRCs (dashed
lines).  The upper set of data and calculations is for $x=1$, while the 
lower are for $x=1.5$. The 6 GeV extension will extend coverage to $Q^2$=8 (10)
GeV$^2$ at $x=1$ (1.5).}
\label{fig:fex}
\end{figure}

Completed and planned measurements~\cite{jlabexpts,e02019} at Jefferson Lab up to 6
GeV will map out the strength of two-nucleon short-range
correlations in nuclei. These measurements will also provide a first look at
high-$Q^2$ cross sections for $2.5 \leq x < 3$, where the cross section is
most sensitive to the presence of multi-nucleon correlations.  
Figure~\ref{fig:fex} shows the structure function per nucleon for iron
compared to a mean-field calculation, a calculation including two-nucleon
SRCs~\cite{ratiocalc}, and one including multi-nucleon SRCs~\cite{multicalc}.
The 4 GeV data are consistent with
two-nucleon SRCs, while future measurements at 6 GeV~\cite{e02019} will 
significantly increase the sensitivity to multi-nucleon correlations,
extending the $Q^2$ range at $x=1.5$ to more than 10 GeV$^2$. Even higher beam
energies will dramatically increase both the $x$ and $Q^2$ range of the data.
With the increased $x$ coverage possible with an 11 GeV beam at Jefferson
Lab, these studies can be extended to isolate and map out the strength of
multi-nucleon correlations.

In addition to improving the sensitivity to multi-nucleon correlations,
higher energies will also provide a new way to study two-nucleon correlations
by allowing us to probe the {\it quark} structure of two-nucleon correlations.
A measurement in the deep inelastic scattering regime at $x>1$ would
allow us to measure for the first time the distributions of superfast quarks
in nuclei, where the struck quark carries more momentum than an entire nucleon.

There is little data available for structure functions of nuclei at $x>1$.
Measurements of muon scattering from iron~\cite{bcdms} have only upper
limits above $x=1.1$, and show a rapid falloff in the structure function near
$x=1$ ($F_2 \propto e^{-16.5x}$).  Measurements of neutrino scattering
from carbon~\cite{ccfr} also have a limited $x$ range ($x<1.2$), but indicate
significantly more strength at large $x$ ($F_2 \propto e^{-8.3x}$).  With the
increase in $Q^2$ possible with the JLab energy upgrade, high precision
measurements of the structure function over a wide range in $x$, allowing us
to cleanly map out the distribution of superfast quarks.  At the $Q^2$ values
achievable with the JLab upgrade, the inelastic cross section dominates even
for $x>1$, and we should be able to extract the structure function in the
scaling regime up to $x \approx 1.4$.

\section{High-density Configurations}\label{sec:high_density}  

These high-$x$ quark distributions are intrinsically related to the quark
structure of the two-nucleon correlations.  Fig.~\ref{fig:fex} demonstrates
that for large $x$ and $Q^2$, the scattering is dominated by scattering from
these short-range configurations.  These quark distributions probe the
internal structure of high-density configurations, where nucleons have
significant overlap. The EMC effect demonstrates that the quark structure of a
nucleus is more than just a convolution of the quark structure of its
nucleons.  If this comes from density-dependent modifications to nucleon
structure, then a similar but much larger effect should be seen when one
examines the quark distributions at large $x$, where the strength comes almost
entirely from nucleon pairs that are nearly overlapping~\cite{sargsian03}.

\begin{figure}[htb]
\vspace*{-.0cm}
                 \insertplot{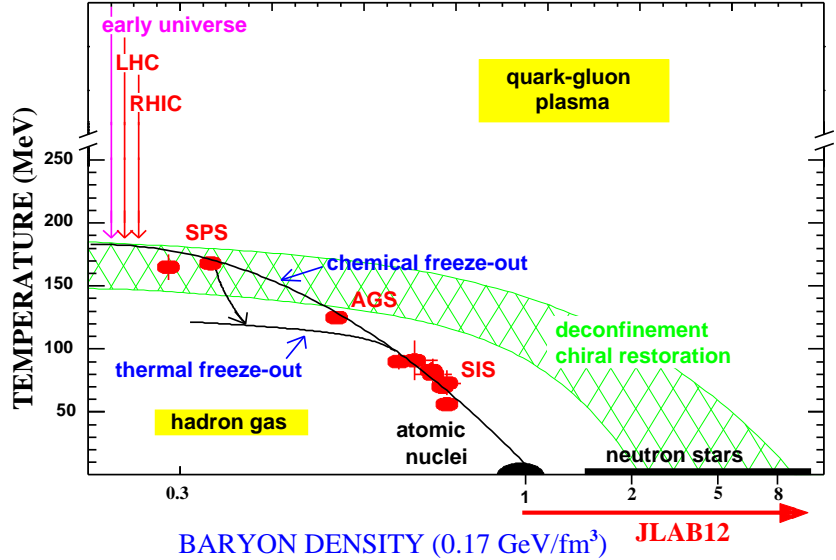}
\vspace*{-0.8cm}
\caption[]{Phase diagram for nuclear matter.}
\label{fig:phase}
\end{figure}

Such a modification of hadron structure is expected at high densities.  Figure
~\ref{fig:phase} shows the phase diagram for nuclear matter.  At high
densities, just as at high temperature, there is a phase transition from
hadronic matter to quark matter.  While ordinary nuclear densities are
well below this transition, the densities achieved in overlapping nucleons
may be high enough to see the effect of this phase transition.  Thus, the
EMC effect may not arise from the increase in {\it average} nuclear density but
instead a be result of large changes in the structure of the high-density
components arising from short-range correlations.

\begin{figure}[htb]
\vspace*{-.0cm}
                 \insertplot{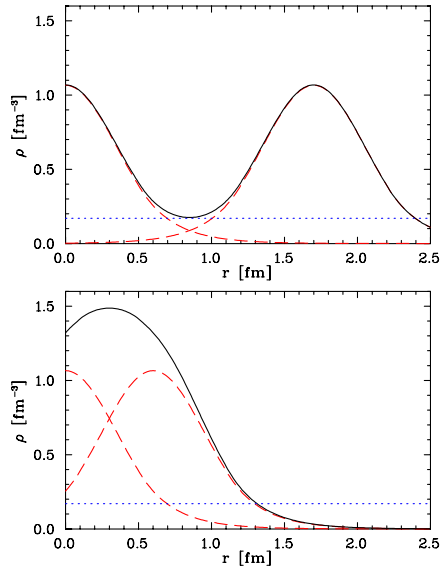}
\vspace*{-0.8cm}
\caption[]{One dimensional density profile for two nucleons separated by 1.7 fm
(top) and 0.6 fm (bottom).}
\label{fig:rho_both}
\end{figure}

Figure~\ref{fig:rho_both} illustrates the overlap of nucleons separated by
1.7 fm, the average separation in nuclear matter, and 0.6 fm.  The short-range
repulsive core of the N-N interaction becomes very large below $\approx$0.4 fm,
so the separation of the nucleons in the correlated pair should be larger than
this distance.  At 0.6 fm, the peak densities are several times normal
nuclear matter densities.

\begin{figure}[htb]
\vspace*{-.0cm}  
                 \insertplot{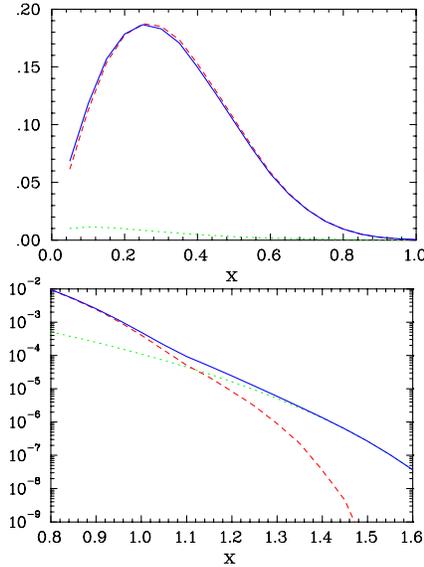}
\vspace*{-0.8cm}
\caption[]{Quark distributions for deuteron assuming only 2-nucleon 
components (dashed line), and assuming a 5\% contribution from a 6-quark bag
(solid line).  The dotted line shows the 5\% 6-quark
component~\cite{mulders84}.}
\label{fig:qofx}
\end{figure}

At these densities, there may be a significant change in the structure of
the nucleon, which will manifest itself in a modified quark distributions.
Figure~\ref{fig:qofx} shows the quark distribution for a deuteron made
of just two nucleons (dashed line), and with a 5\% contribution from a
6-quark bag (solid line).  For $x<1$, there is almost no difference
in the quark distributions, but for $x>1$ the small 6-quark component dominates
the quark distribution.  While this is just one model for nuclear effects
at high density, a significant increase in the distribution of superfast
quarks is a fairly general signature of such modifications. In a nucleon at
rest the quarks share its momentum, and thus cannot have a momentum
fraction, $x$, larger than one.  In nuclei, quarks can gain additional momentum
from the interaction of the nucleons.  Even taking into account the
high-momentum tails of the nucleon distributions coming from SRCs, the quark
distribution falls off extremely rapidly for $x>1$.  However, if the nucleons
in the SRCs have significant overlap, the quarks can directly share their
momentum, allowing for a much greater probability of finding a single quark
with a very large momentum fraction.  Such an observation would be a clear
signature of deviations from the purely hadronic picture of nuclei.

\section{Conclusions}

Measurements of inclusive scattering from nuclei at $x>1$ probe the
high-momentum components of the nuclear wavefunction.  These high momentum
nucleons arise from the short-range interaction of nucleon pairs.
At moderate energy scales, where the scattering is dominated by quasielastic
scattering from individual nucleons, these nucleon pairs
can be kinematically isolated in inclusive reactions.  High energy measurements
will allow us to probe the quark structure of these short-range correlations,
and look for modification to the underlying quark structure of hadrons in
regions where the instantaneous densities are several times the density of
normal nuclear matter.  The structure of nucleons and the distribution of
high-momentum quarks may be substantially altered in this region, where
significant overlap of nucleons should allow direct interaction between quarks
in different nucleons. Studies of these high-density configurations will
allow us to determine if such modifications of nucleon structure are responsible
for the EMC effect, and will help us understand the quark-gluon phase transition
at high density.  This complements RHIC studies of this same transition at high
temperature, while at the same time providing us information on the structure
of matter at extremely high densities.  Probing these high density components
in nuclei is the only way to directly study high density nuclear matter, and
what we learn here will be important in understanding neutron stars and other
compact astronomical objects.

\section*{Acknowledgments}
This work is supported by the U.S. Department of Energy, Nuclear Physics
Division, under contract W-31-109-ENG-38.

\vfill\eject
\end{document}